\documentclass[a4paper,11pt]{article}
\usepackage{pos}
\usepackage{siunitx}

\title{Jets as a probe of the quark--gluon plasma}

\author*[a,b]{Jasmine Brewer}

\affiliation[a]{Center for Theoretical Physics, Massachusetts Institute of Technology,\\
77 Massachusetts Avenue, Cambridge, USA}
\affiliation[b]{Theoretical Physics Department, CERN, CH-1211 Gen\`eve 23, Switzerland}

\emailAdd{jasmine.brewer@cern.ch}

\abstract{
The suppression and modification of high-energy objects, like jets, in heavy-ion collisions provide an important window to access the degrees of freedom of the quark--gluon plasma on different length scales.
Despite increasingly precise and differential measurements of the properties of jets in heavy-ion collisions, however, it has remained challenging to use jets to make unambiguous and model-independent statements about the quark--gluon plasma.
Here I will give a personal take on some origins of these challenges, including the difficulty of modelling and biases from jet selection that obfuscate the direct interpretation of jet modification measurements.
I will discuss a few model studies that have helped to disentangle the source of non-intuitive effects in measurements, and finally highlight data-driven approaches as an interesting opportunity toward studying the quark--gluon plasma in a model-independent way using jets.
}

\FullConference{%
  HardProbes2020\\
  1-6 June 2020\\
  Austin, Texas}

\begin{document}
\maketitle

High-energy collisions between large nuclei provide unique experimental access to a novel high-temperature phase of matter, the quark--gluon plasma, produced in these collisions.
Though the observation of collective behavior of the quark--gluon plasma suggests that it is a strongly-coupled liquid around the scale of its temperature, at high enough momentum-transfer QCD describes quarks and gluons that are weakly-coupled. 
Understanding how a strongly-coupled quark--gluon plasma emerges from QCD, and how its interactions and degrees of freedom interpolate between these regimes, provides strong motivation to study its properties as a function of length scale \cite{Busza:2018rrf}.

An important method for studying the degrees of freedom of a material as a function of length (or energy) scale is to study the attenuation or deflection of an external probe, for example an electron beam, in the material as a function of its energy.
The extremely short lifetime of the quark--gluon plasma makes it impossible to study it with external probes, but high-energy objects produced in the collision can be used instead.
Systematic differences in the yield and properties of high-energy processes in heavy-ion collisions compared to a baseline expectation from proton--proton collisions has become a critical avenue for studying the quark--gluon plasma produced in heavy-ion collisions (see \cite{Connors:2017ptx} for a review of relevant measurements).

In this mini-review, I will give a personal view on the status of using jets to understand the structure of the quark--gluon plasma.
I first mention briefly some types of theoretical frameworks that are employed to study jet modification.
I then discuss challenges in the direct interpretation of jet modification measurements, focusing mostly on biases due to jet selection.
Based on the difficulty of both modelling and the direct interpretation of data, it is highly desirable to constrain the properties of the quark--gluon plasma as independently as possible of the features of specific models.
I will advocate a more agnostic use of models to diagnose the physics effects at play in particular observables, understand their qualitative impact on measurements, and generate new ideas about how to disentangle multiple effects.
I briefly mention a few examples where models are used to understand the origin of non-intuitive features in measurements, and finally discuss some recent effort on data-driven approaches to studying the quark--gluon plasma.

\section{Jets in heavy-ion collisions}

In high-energy collisions between protons, there will occasionally be a particularly high momentum-transfer scattering between partons (quarks or gluons) in the incoming protons. Highly-virtual partons in the final state of this scattering successively fragment to produce collimated sprays of hadrons at the detector, called jets. 
By restricting to those events with jets of high transverse momentum ($p_T$), it is possible to select high momentum-transfer processes for which the QCD coupling is small. Though the full evolution of the parton shower and hadronization is not perturbative, sophisticated machinery has been developed that makes it possible to study jet production in proton--proton collisions in a controlled way using perturbative QCD.

Qualitatively, the goal of reconstructing a jet is to collect all hadrons that originate from radiation off of the same initiating high-energy (``hard'') parton. 
In this ideal case, summing the $p_T$ or invariant mass of all hadrons in the jet gives access to the $p_T$ or invariant mass of the initiating parton.
In practice, asking which hadrons are associated with the hard scattering is not well-defined. At the hadron level a jet must be defined through a ``jet reconstruction algorithm'' that identifies clusters of high-momentum hadrons in the detector. 
Depending on the algorithm, radiation not from the initiating parton that ends up inside the jet cone is counted as part of the jet, and radiation from the initiating parton that is at large angles relative to the jet axis may not be included in the jet.

In heavy-ion collisions, similar high-energy processes can occur between protons in the incoming nuclei. It should be kept in mind that parton distribution functions for nuclei are different than those for protons, but for the production of high energy hadrons these effects seem to be small \cite{ALICE:2012mj, Khachatryan:2016odn, Aad:2016zif}. To the extent that this effect can be neglected for jets, one can think of jets in heavy-ion collisions as being produced as in proton--proton collisions and subsequently modified by the quark--gluon plasma. In this picture, differences in the properties of jets in proton--proton and heavy-ion collisions is a probe of the quark--gluon plasma that jets in heavy-ion collisions must pass through.

The theoretical description of jets in heavy-ion collisions is difficult because many scales of momentum transfer may be relevant for describing their modification in the quark--gluon plasma, including regimes where the QCD coupling is large and perturbative techniques are not fully controlled. 
An energetic parton in a finite-temperature medium will experience radiation induced by the medium in addition to radiation it would have experienced in vacuum. The features of this radiation have been studied extensively (see \cite{Mehtar-Tani:2013pia} for a review) and provide many critical insights on the features of parton interactions with the plasma. To make contact with phenomenology it is however often necessary to treat a jet not as a parton but as a complicated multi-scale object. In this direction there has been substantial effort to formulate Soft Collinear Effective Theory, which has proven to be a powerful tool to study jets in proton--proton collisions, to account for interactions of jets with the medium \cite{Ovanesyan:2011xy, Ovanesyan:2011kn, Vaidya:2020cyi,Vaidya:2020lih}.

A large number of other phenomenological models of jets in heavy-ion collisions are based on Monte Carlo generators for proton--proton collisions (commonly, \textsc{Pythia}) modified to describe interactions with the plasma (for example, \cite{Zapp:2008gi, Casalderrey-Solana:2014bpa, He:2015pra,Tachibana:2017syd, Putschke:2019yrg}).
These descriptions are convenient for making contact with experiment because they define jets through a jet algorithm and can efficiently be used to analyze any observable. 
They can also straightforwardly incorporate nuclear parton distribution functions and a phenomenological model of hadronization.
Though strictly the parton shower is only defined in momentum space, in these models the parton shower is typically assigned a (somewhat ad-hoc) spacetime picture to allow it to be interfaced with the spacetime evolution of the quark--gluon plasma, so that energy lost by partons in the shower can depend on local properties of the medium like temperature or density. In some variants, energy lost by partons from the shower is redistributed in the medium. 
Momentum transferred from the jet to the medium via energy loss means that hadrons ``from the medium'' become correlated with the jet. Experimentally it is necessary to have a procedure for subtracting soft backgrounds that are not correlated with the jet. Especially jet observables that are sensitive to low-energy radiation and radiation further from the jet axis (for example, the jet mass) can be very sensitive to medium response and the treatment of background subtraction \cite{KunnawalkamElayavalli:2017hxo} where many models lose control. Measurements of the distribution of transverse momentum in jets \cite{Khachatryan:2015lha} and, more recently, of large-radius jets \cite{CMS:2019btm,ATLAS:2019rmd} may provide exciting opportunities to learn about this difficult regime from data.
\section{Jet selection and the interpretation of modification observables}
Unfortunately, it is not possible to measure jet-by-jet modification directly, since the properties of a jet when it was produced generally cannot be measured.
The standard procedure is to compare the statistical properties of jets in heavy-ion collisions to a baseline of jets in proton--proton collisions to infer the modification.
This gives rise to several non-trivial issues in the interpretation of jet modification measurements (see also \cite{Busza:2018rrf}).

First, the samples of jets that are selected in proton--proton and heavy-ion measurements are different.
The properties of jets are modified by the quark--gluon plasma, so any method to select heavy-ion jets based on these properties will change the sample compared to the baseline.
Maybe the most important example of this is the selection of jets by their $p_T$: since jets lose energy in the quark--gluon plasma, samples of proton--proton and heavy-ion jets with the same $p_T$ are not comparable because they had systematically different $p_T$ when they were produced.
Each heavy-ion jet was produced with a $p_T$ that was higher by the amount of its energy loss, which itself depends on other aspects of its production.
Almost all properties of jets depend strongly on their $p_T$, so this causes differences in jet observable distributions between proton--proton and heavy-ion collisions.
It is difficult to disentangle this effect from the genuine modification of a jet observable by the quark--gluon plasma that we wish to measure.

In addition, the sample of heavy-ion jets is substantially biased towards those jets that lost as little energy as possible in the quark--gluon plasma.
The reason for this is that the probability for producing a jet falls very steeply with $p_T$.
For a sample of heavy-ion jets with fixed $p_T$, this means that there are many more that were initiated by a hard process that was only slightly higher $p_T$ and lost a small amount of energy than that were produced with much higher $p_T$ and lost a lot of energy, just because the former were produced in dramatically larger numbers due to their lower $p_T$.
This bias can have a particularly large (and confusing) impact on jet modification observables, which will be discussed in more detail in the next Section.

These issues can be reduced in a class of processes where a jet is produced with a high-energy electroweak boson like a photon or $Z$-boson.
Since the (uncolored) electroweak boson is unmodified by the quark--gluon plasma, selecting jets based on the $p_T$ of the recoiling boson eliminates the selection bias present when selecting jets based on their own (modified) $p_T$.
These processes are much more rare than those producing inclusive jets, but have been accessed experimentally \cite{Sirunyan:2017jic,Sirunyan:2017qhf,Aaboud:2018anc,Aad:2020aop}.
The much higher statistics available for inclusive jets, however, provides opportunities for lower uncertainties and more differential measurements, and inclusive jet events additionally have much larger sensitivity to gluon-initiated jets.
It therefore remains critical to use inclusive jet events effectively as a probe of the quark--gluon plasma produced in heavy-ion collisions.
To do this requires thinking carefully about how to disentangle physical effects from selection biases.

\section{Disentangling competing effects in models}

With substantial contributions from selection biases that complicate the interpretation of jet modification measurements, models can play an important role in diagnosing the physics that a measurement is sensitive to.

Jet selection biases can qualitatively change the interpretation of jet modification observables.
An important example is understanding whether jets get narrower or wider due to their interaction with the plasma.
The measured jet shape in heavy-ion collisions \cite{Chatrchyan:2013kwa,Sirunyan:2018jqr} is narrower at intermediate radii than for jets in proton--proton collisions in the same $p_T$ range.
However, this narrowing can be understood as an increase in the fraction of (typically narrower) quark-initiated jets in heavy-ion collisions due to the larger energy loss of gluon-initiated jets  \cite{Chien:2015hda}.
It was shown in \cite{Rajagopal:2016uip} and explored in more detail in \cite{Brewer:2017fqy} that a narrowing of the jet shape can occur even if every individual jet widens due to its interaction with the plasma.
This narrowing is not observed in the measurement of photon-tagged jet shapes \cite{Sirunyan:2018ncy} where the selection bias in $p_T$ is removed.
It has also been argued that a similar effect may impact fragmentation functions \cite{Caucal:2020xad} and that selection biases may also substantially decrease the sensitivity of jet observables to medium response \cite{Brewer:2020chg}.

Another unexpected realization, first suggested in \cite{Milhano:2015mng}, is that the energy asymmetry of back-to-back jets may not be due to differing path lengths of each jet in the quark--gluon plasma.
Non-intuitively, samples of jets in this model with the same path length have similar asymmetry to those with varying path lengths, suggesting that fluctuations in the energy loss of jets with the same path length could be crucial to the asymmetry.
The fact that energy loss fluctuations can be comparably important to path-length differences has also been observed in \cite{Escobedo:2016jbm}, and \cite{Brewer:2018mpk} found a delicate interplay between the path-length dependence and the role of fluctuations.
Whether this originates from path-length effects being much smaller jet-by-jet than energy loss fluctuations, or due to selection biases, remains to be understood.

Finally, it has been observed that the jet $R_{AA}$ doesn't depend on collision energy, despite the larger energy loss anticipated in the hotter plasma produced at higher energies.
In \cite{He:2018xjv}, this is due to simultaneous effects of different spectra and different energy loss at the two energies.
This reinforces the sensitivity of the jet $R_{AA}$ to the shape of the vacuum spectra, and thus highlights the importance of measuring energy loss more directly, for example \cite{Brewer:2018dfs}.

\section{Toward interpreting data without models}

Due to the challenges posed by selection biases and the lack of first-principles modelling, there has been substantial recent work towards increasingly model-agnostic and data-driven approaches to learning about the quark--gluon plasma.

In \cite{Brewer:2018dfs} a method was proposed to reduce the effects of selection biases in jet modification observables by adjusting the $p_T$ range of a heavy-ion jet measurement relative to the proton--proton baseline to compensate for the average energy loss.
The average energy loss of jets as a function of their $p_T$ can be measured from the ($p_T$-dependent) horizontal shift of the jet-production spectrum in heavy-ion collisions compared to proton--proton collisions \cite{Brewer:2018dfs}.
This procedure only corrects for the average energy loss, so the samples of heavy-ion and proton--proton jets that are compared are still not identical.
However, it reduces the impact of selection biases, and additionally provides a benchmark for the size of selection bias effects in jet modification observables.
It would additionally be interesting to compare this to using a Bayesian analysis to extract the average energy loss from fits to $R_{AA}$ for inclusive and photon-tagged jet measurements as in \cite{He:2018gks}.
A potentially exciting area for future exploration is understanding how boson+jet and hadron+jet measurements can be used in tandem with inclusive jet measurements to clarify their interpretation \cite{Casalderrey-Solana:2018wrw}.

The difference in energy loss between jets initiated by quarks or gluons provide a theoretically clean probe of the interactions of the quark--gluon plasma with jets.
To the extent that the quark--gluon plasma sees a jet as an extended object with the color of the initiating parton, gluon jets will lose more energy by a factor of their larger color charge.
Deviations from this scaling would suggest that the quark--gluon plasma resolves finer scales within the jet \cite{Apolinario:2020nyw}.
Unfortunately, jet measurements are generally a mixture of both quark- and gluon-initiated jets, which makes constraining their separate energy loss difficult.
The quenching of jets initiated by a b-quark has been measured \cite{Chatrchyan:2013exa,Sirunyan:2018jju} and shows no significant difference compared to the quenching of inclusive jets, which also contain gluon-initiated jets.
Particularly at high $p_T$ in heavy-ion collisions however, inclusive jets may have quite high quark fraction, which could make this comparison relatively insensitive to gluon energy loss.
There has been recent work aiming to use a Bayesian analysis to extract $R_{AA}$ separately for quark and gluon jets from experimental data \cite{Qiu:2019sfj} which found large differences in the quenching of quark- and gluon-initiated jets.
A recent method for extracting quark and gluon fractions from data using \textsc{Pythia} templates, on the hand, has suggested minimal differences between proton--proton and heavy-ion collisions \cite{Sirunyan:2020qvi}.
A method for extracting quark and gluon fractions without using templates has recently been proposed \cite{Metodiev:2018ftz} and applied to heavy-ion collisions \cite{Brewer:2020och} and may enable future data-driven extraction of quark and gluon jet modification.

Exciting recent work has also highlighted how future measurements can provide novel handles on energy loss and modification.
For example, it was proposed in \cite{Apolinario:2017sob} that the intermediate $W$ boson in the decay of a top quark makes it possible to change the amount of time a jet interacts with the quark--gluon plasma, and could be accessed experimentally at the Future Circular Collider.
Another crucial open question is whether there is jet quenching in small collision systems, such as proton--lead, where smaller expected energy loss necessitates higher precision.
It was proposed recently that minimum-bias oxygen--oxygen collisions can substantially reduce uncertainties by eliminating centrality selection, and thus provide an important opportunity for discovering energy loss in small systems \cite{Huss:2020dwe,Huss:2020whe}.

\section{Concluding remarks and outlook}

Recent years have seen dramatic progress in increasingly differential measurements of a variety of jet observables in heavy-ion collisions.
These provide an exciting opportunity to access the short-distance structure of the quark--gluon plasma.
In tandem with theoretical developments and improvements in phenomenological modelling, model-agnostic and data-driven approaches provide an important, and complementary, opportunity to cross-check model assumptions and draw model-independent conclusions where possible.

\section{Acknowledgements}
I would like to thank the Hard Probes 2020 Organizing Committee for the invitation to give an overview of the medium modification of jets in heavy-ion collisions. I gratefully acknowledge valuable discussions with Quinn Brodsky, Yang-Ting Chien, Eliane Epple, Raghav Kunnawalkam Elayavalli, Gian Michele Innocenti, Vit Kucera, Yen-Jie Lee, Aleksas Mazeliauskas, Guilherme Milhano, Lina Necib, Krishna Rajagopal, Jesse Thaler, Urs Wiedemann, Xiaojun Yao, Yi Yin, and Nima Zardoshti.

\bibliographystyle{JHEP}
\bibliography{refs}

\end{document}